\documentclass[12pt,twoside]{article}
\usepackage{fleqn,espcrc1}

\input{psfig}
\usepackage{graphicx}

\usepackage[figuresright]{rotating}

\def\Journal#1#2#3#4{{#1} {\bf#2} (#4) #3}

\def\NPA{{\rm Nucl. Phys.} A}
\def\NPB{{\rm Nucl. Phys.} B}
\def\PLB{{\rm Phys. Lett.}  B}

\def\PRD{{\rm Phys. Rev.} D}

\def\la{\langle}
\def\ra{\rangle}

\def\be{\begin{equation}}
\def\ee{\end{equation}}
\def\bea{\begin{eqnarray}}
\def\eea{\end{eqnarray}}

\newcommand{\AmS}{{\protect\the\textfont2
  A\kern-.1667em\lower.5ex\hbox{M}\kern-.125emS}}

\begin{document}
\title{ Exploring timelike exclusive processes in the light-front approach
}

\author{Chueng-Ryong Ji\address[MCSD]{Department of Physics,
        North Carolina State University\\
        Raleigh, NC 27695-8202, USA}%
        and 
        Ho-Meoyng Choi\addressmark[MCSD]
         \address{Department of Physics, Carnegie Mellon University\\ 
           Pittsburgh, PA 15213, USA}
        }


\maketitle

\begin{abstract}
We discuss a necessary nonvalence contribution in timelike exclusive 
processes. Utilizing a Schwinger-Dyson type of approach, we relate the 
nonvalence contribution to an ordinary light-front wave function
that has been extensively tested in the spacelike exclusive processes.
An application to $K_{\ell3}$ decays provides encouraging results.
\end{abstract}

\baselineskip=20pt
Not only the new and upgraded B-meson factories but also the lower-lying
meson facilities such as the $\tau$-Charm factories at Cornell and the
recent PEP-N project at SLAC demand intensive theoretical analyses of
exclusive meson decays and form factors. Unlike the leading twist 
structure functions measured in deep inelastic scattering, such exclusive
channels are sensitive to the structure of the hadrons at the 
amplitude level and to the coherence between the contributions of the 
various quark currents and multi-parton amplitudes. The central unknown 
required for reliable calculations of weak decay amplitudes are thus
the hadronic matrix elements. 

Perhaps, one of the most popular formulations for the analysis of
exclusive processes involving hadrons may be provided in the framework of 
light-front (LF) quantization~\cite{BPP}. 
In particular, the Drell-Yan-West ($q^+=q^0+q^3=0$) frame 
has been extensively used in the calculation
of various electroweak form factors and decay 
processes~\cite{CJ1,Kaon}. As an example,
only the parton-number-conserving (valence) Fock state contribution 
is needed in $q^+=0$ frame when the ``good" component of the current,
$J^+$ or ${\bf J}_{\perp}=(J_x,J_y)$, is used for the spacelike 
electromagnetic form factor calculation of pseudoscalar mesons\cite{zero}.
On the other hand, the analysis of timelike exclusive processes 
has remained
as a rather significant challenge in the LF approach. In principle, the 
$q^+\neq0$ frame can be used to compute the timelike processes but 
then it is inevitable to encounter the particle-number-nonconserving 
Fock state (or nonvalence) contribution.
The main source of difficulty in constituent quark model(CQM)
phenomenology is the lack of information on the non-wave-function 
vertex(black blob 
in Fig.~1(a)) in the nonvalence diagram arising from the quark-antiquark 
pair creation/annihilation. 

\begin{figure}[t]
\centerline{\psfig{figure=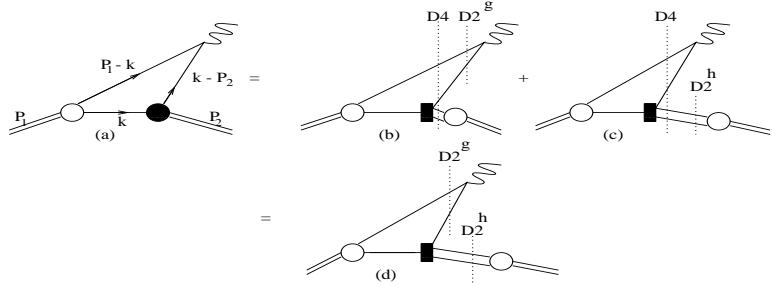,height=1.5in,width=4.0in}}
\caption{Effective treatment of the LF nonvalence amplitude.}
\end{figure}

In this talk, we thus present a way of handling the nonvalence contribution. 
Our aim of new treatment is to make the program more
suitable for the CQM phenomenology specific to the low momentum transfer 
processes. More details of our effective treatment can be found 
in Ref.~\cite{JC}.

\begin{figure}[t]
\centerline{\psfig{figure=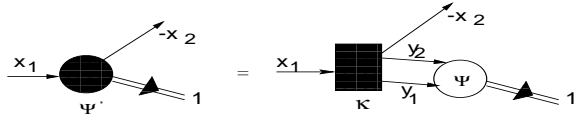,height=0.6in,width=3.0in}}
\caption{Non-wave-function vertex(black blob) linked to an ordinary
LF wave function(white blob).}
\end{figure}

The crux of our method~\cite{JC} is the link between the non-wave-function
vertex (black blob) and the ordinary LF wave function (white blob)
as shown in Fig.~2:
\bea\label{eq:SD}
(M^2-M'^{2}_0)\Psi'(x_i,{\bf k}_{\perp i})
=\int[dy][d^2{\bf l}_{\perp}]
{\cal K}(x_i,{\bf k}_{\perp i};y_j,{\bf l}_{\perp j})
\Psi(y_j,{\bf l}_{\perp j}),
\eea 
where $M$ is the mass of outgoing meson and $M'^{2}_0=(m^2_1+{\bf 
k}^2_{\perp 1})/x_1 - (m^2_2+{\bf k}^2_{\perp 2})/(-x_2)$ with
$x_1 = 1-x_2 > 1$ due to the kinematics of the non-wave-function vertex.
With this link made by a Schwinger-Dyson(SD) type equation  
(Eq.~(\ref{eq:SD})), we can now get the nonvalence contribution (Fig.1(a))
as a sum of LF time-ordered amplitudes (Figs.1(b) and (c)) and moreover
find that the four-body energy denominator $D_4$ is exactly cancelled in 
summing the LF time-ordered amplitudes; i.e.,
$1/D_4D^g_2 + 1/D_4D^h_2 =1/D^g_2D^h_2$.
We thus obtain the amplitude identical to the nonvalence
contribution in terms of ordinary LF wave functions of gauge boson($W$)
and hadron (white blob) as drawn in Fig.1(d). This method, however,
requires to have some relevant operator depicted as the black
square(${\cal K}$) in Fig.~2(See also Fig.1(d)), that is in general
dependent on the involved momenta connecting one-body to three-body sector.
While the relevant operator ${\cal K}$ is in general
dependent on all internal momenta ($x,{\bf k}_{\perp},y,{\bf l}_{\perp}$),
a sort of average on ${\cal K}$ over $y$ and ${\bf l}_{\perp}$
depends only on $x$ and ${\bf k}_{\perp}$(See Eq.~(\ref{eq:SD})).
In the semileptonic decay processes involving small momentum transfers
such as the $K_{\ell3}$ decays, we can kinematically justify~\cite{JC} that
the r.h.s. of Eq.~(\ref{eq:SD}) may be approximated as a constant.

If the initial and final mesons are pseudoscalar ($0^{-+}$), the relevant
matrix element for the semileptonic decay of the initial meson 
($Q_{1}\bar{q}$ bound state) with four-momentum $P^\mu_1$ and mass $M_1$ 
into the final meson ($Q_{2}\bar{q}$ bound state) with 
$P^\mu_2$ and $M_2$ is given by
\be{\label{eq:K1}}
J^{\mu}(0)=\la P_{2}|\bar{Q_{2}}\gamma^{\mu}Q_{1}|P_{1}\ra
= f_{+}(q^{2})(P_{1}+P_{2})^{\mu} + f_{-}(q^{2})q^{\mu},
\ee
where $q^\mu=(P_{1}-P_{2})^\mu$ is the four-momentum transfer to
the lepton pair ($\ell\nu$) and $m^{2}_\ell\leq q^2\leq (M_{1}-M_{2})^{2}$.
We compute~\cite{JC} the matrix element in a purely longitudinal momentum
frame where $q^+>0$ and ${\bf P}_{1\perp}={\bf P}_{2\perp}=0$ so that
$q^2=q^+q^->0$. For the check of frame-independence,
we also compute the ``$+$" component of the current $J^\mu_{D}$
in the Drell-Yan-West ($q^+=0$) frame where only valence contribution
exists. Since the form factor $f_+(q^2)$ obtained
from $J^+_D$ in $q^+=0$ frame is immune to the zero-mode
contribution~\cite{zero},
the comparison of $f_+(q^2)$ in the two completely different frames
(i.e. $q^+=0$ and $q^+\neq0$) would reveal the validity of existing model
with respect to a covariance(or frame-independence).
The comparison of $f_-(q^2)$, however, cannot give a meaningful test
of covariance because of the zero-mode complication as noted in
Ref.~\cite{zero}. Indeed, the difference between the two ($q^+=0$ and
$q^+\neq0$) results of $f_-(q^2)$ amounts to the zero-mode
contribution.

In our numerical calculation for the processes of $K_{\ell3}$ decays, 
we use the linear potential parameters
presented in Refs.~\cite{CJ1,Kaon}.
In Table~1, we summarize the experimental observables for the
$K_{\ell3}$ decays, where $\lambda_i=M^{2}_{\pi}f'_{i}(0)/f_{i}(0)(i=+,0)$
and $\xi_{A}=f_-(0)/f_+(0)$.
Incidentally, the $K_{\ell3}(\ell=e,\mu)$ decays involving rather 
low momentum transfers 
bear a substantial contribution from the nonvalence part and their 
experimental data are better known than other semileptonic processes with 
large momentum transfers.
As one can see in Table~1, our new results (column
2) are now much improved and comparable with the data.
More results including heavier mesons are discussed in Ref.~\cite{JC}.
\begin{table}[t]
\caption{ Model predictions for the parameters of $K^{0}_{\ell3}$ decays.
The decay width is in units of $10^{6}$ s$^{-1}$. The
used CKM matrix is $|V_{us}|=0.2196\pm0.0023$ from the Particle Data Group, 
D. E. Groom {\em et al.}, Eur. Phys. J.  C {\bf 15}, 1 (2000).
}
\begin{center}
\begin{tabular}{|c|c|c|c|}
\hline
&Effective & $q^+=0$ & Experiment\\
\hline
$f_{+}(0)$ & 0.962 [0.962] & 0.962 [0.962] & \\
\hline
$\lambda_{+}$& 0.026 [0.083] & 0.026 [0.026] & $0.0288\pm0.0015
[K^{0}_{e3}]$\\
\hline
$\lambda_{0}$& 0.025 [$-0.017$]& 0.001 [$-0.009$]
& $0.025\pm0.006[K^{0}_{\mu3}]$\\
\hline
$\xi_{A}$& $-0.013$ [$-1.10$]& $-0.29 [-0.41]$
& $-0.11\pm0.09[K^{0}_{\mu3}]$\\
\hline
$\Gamma(K^{0}_{e3})$ & $7.3\pm0.15$
&  $7.3\pm0.15$ & 7.5$\pm$0.08\\
\hline
$\Gamma(K^{0}_{\mu3})$ & $4.92\pm0.10$
&  $4.66\pm0.10$ & 5.25$\pm$0.07\\
\hline
\end{tabular}
\end{center}
\end{table}

In summary, we presented an effective treatment of the LF
nonvalence contributions crucial in the timelike exclusive processes.
Using a SD-type approach and summing the LF time-ordered amplitudes,  
we obtained the nonvalence contributions in terms of ordinary LF 
wave functions of gauge boson and hadron
that have been extensively tested in the spacelike exclusive processes
\cite{CJ1,Kaon}.  
Including the nonvalence contribution, our results on 
$K_{\ell3}$ not only show a definite 
improvement in comparison with experimental data but also exhibit a 
covariance (i.e frame-independence) of our approach. 

This work was supported by the US DOE under contracts DE-FG02-96ER40947.
The North Carolina Supercomputing Center and the National Energy Research
Scientific Computer Center are also acknowledged for the grant of 
supercomputer time.

\end{document}